# LPKI - A Lightweight Public Key Infrastructure for the Mobile Environments[†]

Mohsen Toorani [‡]

Ali A. Beheshti

*Abstract*— **The non-repudiation as an essential requirement of many applications can be provided by the asymmetric key model. With the evolution of new applications such as mobile commerce, it is essential to provide secure and efficient solutions for the mobile environments. The traditional public key cryptography involves huge computational costs and is not so suitable for the resource-constrained platforms. The elliptic curve-based approaches as the newer solutions require certain considerations that are not taken into account in the traditional public key infrastructures. The main contribution of this paper is to introduce a Lightweight Public Key Infrastructure (LPKI) for the constrained platforms such as mobile phones. It takes advantages of elliptic curve cryptography and signcryption to decrease the computational costs and communication overheads, and adapting to the constraints. All the computational costs of required validations can be eliminated from end-entities by introduction of a validation authority to the introduced infrastructure and delegating validations to such a component. LPKI is so suitable for mobile environments and for applications such as mobile commerce where the security is the great concern.**

## I. INTRODUCTION

Nowadays, the *Public Key Cryptography* (PKC) is an essential security mechanism for any open and popular system. The conventional public key algorithms are usually slow and involve huge computational costs so they are not so suitable for the mobile environments and resource-constrained devices that are dealing with memory, power, and computational restrictions. With the evolution of new applications such as mobile commerce, deploying PKC by the mobile devices is unavoidable. The non-repudiation that is an essential security service for any application with the monetary value can be provided by the PKC.

Although the processing capabilities of mobile phones are increasingly enhanced, and the newer mobile phones perform live video stream processing, designing secure and efficient enough cryptographic techniques is still an important issue since it can reduce the power consumptions and processing time. It is also an essential task especially when the cryptographic computations should take place in smart cards. While the PKC is generally used for security establishment at the application layer, it can be a suitable solution for the lower layers if the computational costs are lightened.

The *Elliptic Curve Cryptography* (ECC) is usually deemed as a suitable solution for the resource-constrained devices [1]. However, the conventional *Public Key Infrastructures* (PKI) such as PKIX and *Wireless Public Key Infrastructure* (WPKI) are usually based on modular exponentiation (RSA) that is not so suitable for the resource-restricted platforms. This has stimulated many efforts to use ECC in PKIs [2, 3]. In EC-based systems, the public keys are obtained using the ECC, the signatures are usually EC-based, and certain considerations should be taken into account that are not the case in the traditional PKIs.

In this paper, a Lightweight Public Key Infrastructure for the mobile environments, called LPKI, is briefly introduced that has considerably light computational costs since it is based on ECC and uses signcryption. LPKI introduces several improvements to the conventional PKIX to efficiently decrease its computational costs and make it suitable for the resource-constrained platforms. This paper is organized as follows. Some preliminaries on ECC and signcryption are provided in Sections 2 and 3 respectively. The proposed infrastructure (LPKI) and its components are briefly introduced in Section 4, and Section 5 gives the conclusions.

## II. ELLIPTIC CURVE CRYPTOGRAPHY

Currently, the ECC has revolutionized the arena of security establishment. The EC-based solutions are usually based on difficulty of solving the *Elliptic Curve Discrete Logarithm Problem* (ECDLP) and factorization in elliptic curves [4]. The EC-based systems can attain to a desired security level with significantly smaller keys than that of required by their exponentiation-based counterparts. As an example, it is believed that a 160-bit key in an elliptic curve-based system provides the same level of security as that of a 1024-bit key in an RSA-based system [4]. This creates great efficiencies in key storage, certificate size, memory usage, and required processing so it enhances the speed and leads to efficient use of power, bandwidth, and storage that are the basic limitations of resource-constrained devices.

The EC-based approaches have a great computational advantage over their modular exponentiation-based counterparts when they are executed on constrained platforms [1]. Table I illustrates such an advantage for the running-time of ECDSA and RSA digital signatures over two mobile phones.

TABLE I. A COMPARISON BETWEEN THE EXECUTION TIMES (IN MS) [5]

| Mobile Phone | Signature generation | | Signature verification | |
|---|---|---|---|---|
| | *ECDSA* | *RSA* | *ECDSA* | *RSA* |
| Nokia 6610 | 2.294 | 74.682 | 4.382 | 2.825 |
| Siemens S55 | 18.963 | 883.602 | 35.277 | 30.094 |

## III. SIGNCRYPTION

The signcryption is a relatively new cryptographic technique that is supposed to fulfill the functionalities of digital signature and encryption in a single logical step, and can effectively decrease the computational costs and communication overheads in comparison with the traditional signature-then-encryption schemes. In the traditional signature-then-encryption schemes, a message is digitally signed then followed by an encryption that has two problems: Low efficiency and high cost of such summation, and the case that any arbitrary scheme cannot guarantee the security. The first signcryption scheme was introduced by Zheng in 1997 [6]. Zheng also proposed an EC-based signcryption scheme that saves 58% of computational and 40% of communication costs when compared with the traditional EC-based signature-then-encryption schemes [7]. Several signcryption schemes are also proposed during the years, each of them offering different level of security services and computational costs. The concept of signcryption and unsigncryption is depicted in Fig. 1 and 2 respectively.

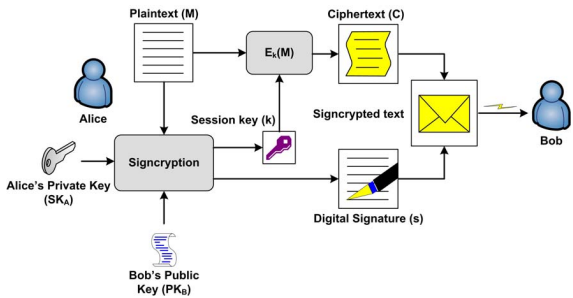

Figure 1.  Signcryption

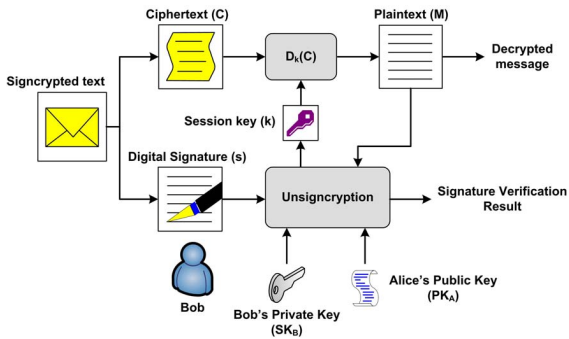

Figure 2.  Unsigncryption

## IV. THE PROPOSED INFRASTRUCTURE (LPKI)

The PKI is a set of hardware, software, people, policies, and procedures needed to create, manage, store, distribute, and revoke digital certificates [8]. Since description of a PKI takes hundreds of pages and due to the space limitations, we do not describe the common points that can be easily found in the PKIX references. Due to the popularity of X.509 certificates and its well-tried infrastructure (PKIX), we define our proposed infrastructure (LPKI) to have as much as similarities to the PKIX. The main differences between our proposed solution and the popular infrastructures such as PKIX and WPKI are as follows:

1) Using ECC. This is opposed to the traditional, well-tried, and currently available infrastructures that are based on the modular exponentiation (RSA).

2) Using symmetric encryption in which the PKC is used for the session key establishment. This is opposed to many of currently available schemes that use asymmetric algorithms for encryption.

3) Using signcryption whenever both of encryption and digital signature are simultaneously required. This is opposed to the traditional systems deploying signature-then-encryption schemes.

4) Assigning only one pair of private-public keys to each subscriber. This is opposed to some infrastructures such as WPKI that devotes two pairs of private-public keys to each subscriber, one pair for encryption, and the other pair for digital signature.

5) Introducing a new component called *Validation Authority* (VA) for performing the required validations for the end-entities.

6) Taking into account several considerations such as public key validation that are essential in EC-based systems and are not considered in the traditional PKIs.

Such improvements can effectively decrease the computational costs and communication overheads. Since LPKI is based on the ECC, it takes into account several considerations that are not the case in the traditional exponentiation-based infrastructures. The remaining of this paper is devoted to the distinguished specifications of the LPKI whereby it differs from its counterparts. This includes its components, public/private key and certificate generation, certificate repository, and the preferred security mechanisms. A typical low-level specification on domain parameters and certain considerations on their selections can be found in [9]. The remained issues correspond with the PKIX specifications.

### A. Components

LPKI consists of the following components:

- *Registration Authority* (RA): Takes care of registering the subscriber's details and issuing a unique identifier for each subscriber.

- *Certification Authority* (CA): Takes care of issuing and managing certificates.

- Digital certificates: Strings of information that binds the unique identifier of each subscriber to his/her corresponding public key.

- *Certificate Repository* (CR): Takes care of certificate repository.
- *OCSP Server*: Takes care of responding to the inquiries about the certificates' revocation statuses. It may be deemed as a part of CA.
- *Validation Authority* (VA): A *Trusted Third Party* (TTP) that accomplishes all the required validations for the end-entities.
- *Key Generating Server* (KGS): An optional component that takes care of generating public-private keys and may be deemed as a part of RA. It is not required if the security policies allows end-entities to generate their own public-private keys and they have adequate capabilities to do this.
- End-entities: Mobile Equipments (ME) such as mobile phones and PDA with the following duties:
  - Generating the subscriber's public-private key pair, Applying for the first-time certification issuance, Requesting for the key pair update and certificate-renewal (optional),
  - Storing and allowing access to the subscriber's public-private key pair,
  - Retrieving a certificate and its revocation status, Performing the required validations including certificate validation and static/ephemeral public key validations (optional),
  - Taking care of security mechanisms.
- Timestamp Server: Produces unique and valid timing information for the LPKI's components.

*B. Public / Private Key Generation*

For each user $U$, the private key is a randomly selected integer $SK_U \in_R [1, n-1]$, and the public key is a point of elliptic curve that is generated as $PK_U = SK_U G$. Each subscriber is identified with a unique identifier ($ID_U$). For the case of mobile networks, the unique identifier can be the subscriber's international phone number, e.g. MSISDN in the GSM. LPKI supports two modes of public key generation:

*Mode 1:* Generating the public-private keys in the KGS, as it is depicted in Fig. 3.

*Mode 2:* Generating the public-private keys at the end-entities, as it is depicted in Fig. 4.

In the first mode, the public-private keys are generated in the trusted KGS. The generated public keys are then certified by the CA and the appropriate information is directly and securely stored on the subscriber's smart card, e.g. the SIM, as it is depicted in Fig. 3. According to the security policies, the CA may store the private key and other relevant information in a secure module as the key backup center, in order to use it for the future key recoveries.

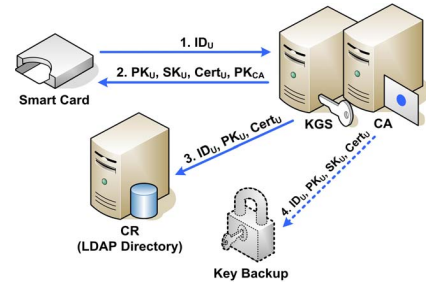

Figure 3. Public-Private Key generation in the KGS (mode 1)

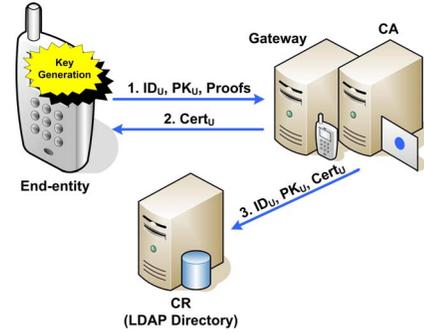

Figure 4. Public-Private Key generation at end-entities (mode 2)

In the second mode, the keys are generated at the end-entities, and they are stored on their smart cards or other security modules. The CA will then certify the generated public keys after adequate proofs. The CA should verify that the end-entities really possess the corresponding private keys of their claimed public keys. This may be accomplished by a zero-knowledge technique. Since resource-constrained devices are not capable of generating very strong random numbers to be used as private keys, and due to the inherent vulnerabilities of wireless interface, the former approach has an obvious security advantage over the latter. However, the second mode provides more flexibility since the procedures of automatic key update, and certificate renewal can be easily accomplished.

Several considerations should be taken into account when the public keys are to be generated at end-entities (mode 2). Traditionally, the public key validation is not considered in the PKIX standards (such as [10] and [11]), and the CA performs just a proof of possession by checking the user's signature over a message of a predetermined format. This can be misused for an *invalid-curve attack* [4] when deploying the ECC. In the EC-based systems, the CA not only should perform the proof of possession but should also perform the public key validation. Otherwise, the CA may certify an invalid public key that is a key of a small order resided on an invalid-curve. Antipa et al. [12] demonstrated how to get a certificate for an invalid public key when the CA uses the ECDSA for digital signature. To thwart the *invalid-curve* attack, the CA should perform the public key validation. The public key of user $U$, $PK_U = (x, y)$ is valid if all the following conditions are simultaneously satisfied [12]:

(a) $PK_U \neq O$ where $O$ denotes the point of elliptic curve at infinity.

(b) $x$ and $y$ should have the proper format of $F_q$ elements.

(c) $PK_U$ should satisfy the defining equation of $E$.

The end-entities should have access to an authentic copy of the CA's public key, in order to use it for verifying the CA's signatures. For the case of mobile systems that use the SIM, the private key can be encrypted using the subscriber's password or PIN (Personal Identification Number), and it can be stored in an elementary file of the SIM. The hash value of the private key that can be used for validation of the encrypting password can also be stored in the SIM. The public keys can be stored in a transparent elementary file. For the case of GSM and according to the GSM 11.11 standard, the trusted keys and certificates are stored in 4FXX files, and the key information can be stored in file 4F50.

*C. Digital Certificates*

Each subscriber should get a digital certificate $Cert_U$ for his/her public key from the CA. In the LPKI, each subscriber's identifier is associated with only one pair of public-private keys so the CA issues only one certificate for each subscriber. Unlike the WPKI that has reduced the required amount of certificate storage by omitting some fields from the X.509 certificate's format, the LPKI uses the same format of the popular and well-tried X.509v3 certificates. However, all the signatures over the certificates should be EC-based with the same domain parameters of the LPKI. With the current evolution of technology, both smart cards (e.g. the SIM) and handheld devices have adequate storage capacity that they do not have any problem with the format of X.509 certificates, especially when each subscriber is associated with only one digital certificate in the LPKI. Since all the required validations will be delegated to the VA, the involved entities are not dealing with the individual certificates so their resource constraints will not cause any problem with the format of X.509 certificates. Since LPKI uses the same format of X.509v3 certificates, it has a compact compatibility with the LPKI, and its certificates can be used in any other compatible infrastructure and system.

*D. Certificate Repository*

For each subscriber, a set of information including the unique identifier, the public key, and the issued certificate are stored in the CR, as it is depicted in Fig. 3 and 4. The CR in the LPKI stores certificates in an LDAP (Lightweight Directory Access Protocol) directory [13]. The LDAP [14] that was initially developed for accessing, querying, and manipulating the X.500-based directory services on the Internet, supports TCP/IP, uses a hierarchical tree-like structure, and represents a lightweight, fast, and scalable directory service. The LDAP records must be updated at the appropriate times, and when a certificate is issued or revoked.

*E. Security Mechanisms*

The confidentiality, authentication, integrity, and non-repudiation are the most desired security services for many systems that are provided with the security mechanisms such as encryption and digital signature. In the LPKI, all the messages are encrypted using a secure symmetric encryption algorithm such as AES where its secret key is established using the public-private keys of the involved entities. Although any authenticated key exchange protocol can be deployed, LPKI takes advantage of EC-based and improved version of the HMQV key exchange protocol [15]. This is because of its efficiency, standardization, and several number of provided security attributes. LPKI provides only one pair of public-private keys for both encryption and digital signature. This can be accomplished using the signcryption schemes that need to only one pair of keys for simultaneously performing encryption and digital signature. Some signcryption schemes can be made generalized, i.e. it is possible to use them for the purpose of encryption, digital signature, or both of them. If the key backup and key recovery is not considered in the security policies, it may be necessary to choose a signcryption scheme that provides the attribute of public verifiability so the judge can easily verify the previously issued signatures of end-entities without any need for their private keys, even if the end-entities have lost their own private keys.

*F. Certificate Validation and Messages flows*

Any public key is not trustworthy unless it is verified using its corresponding validated certificate. The certificate validation mainly includes [16]:

1. Verifying the integrity and authenticity of the certificate by verifying the CA's signature over it.

2. Verifying that the certificate is not expired.

3. Verifying that the certificate is not revoked.

In order to perform the certificate validation, it is necessary to obtain the revocation status of the certificate. Generally, the revocation status can be obtained using either of *Certificate Revocation List* (CRL), Delta CRL, or *Online Certificate Status Protocol* (OCSP). Since OCSP [17] is the most profitable solution for the resource-constrained devices that are not capable of saving a too large CRL, and according to its great advantage due to its online and timely responses, LPKI has introduced the "*OCSP Server*" as one of its components. The OCSP responses are digitally signed with a private key that its corresponding trusted public key is known to the end-entities.

Two modes of information retrieval and message exchange are predicted in the LPKI, as are depicted in Fig. 5 and 6 in which $OCSP_A$ and $OCSP_B$ denote the corresponding OCSP tokens for the certificates of *Alice* and *Bob*. In mode "1", it is assumed that the end-entities are capable of performing the certificate validation, and static/ephemeral public key validations. In mode "2", on the other hand, all validations including the certificate validation and ephemeral/static public key validations of both end-entities are delegated to the VA so it offers significant advantages for the resource-constrained platforms due to its decreased computational costs and communication overheads. This is because the required time of sending a certificate to a validation server and subsequently, receiving and authenticating the response can be considerably less than the required time of performing the necessary path discovery and validations on a resource-constrained device.

In mode "1", both of end-entities should individually perform the certificate and static/ephemeral public key validations. However, in mode "2", all the signcrypted messages are directed to the VA through the communication link, and the VA accomplishes all the required validations for both of sender and recipient end-entities. The VA will contact the designated entity after a successful validation. If any error occurs, the VA will send an error message to the initiator and saves a copy in its log file. According to the security policies, all the transmitted messages may be separately saved by the VA in a storage media.

The VA accomplishes the certificate validation via the *Delegated Path Validation* (DPV) protocol [18]. The VA queries CR for the certificates of both sender and recipient through their identifiers. It digitally signs its responses unless an error is occurred. A hash value of all the transmitted parameters in addition to the identifiers of both sender and recipient end-entities should be included in its signed responses. The VA obtains its required revocation statuses by getting OCSP responses from the OCSP server.

If an end-entity needs to know the public key or revocation status of another end-entity, according to its desired mode of operation, it sends the identifier of the recipient end-entity in addition to either of tags "1" or "2" that indicates the requested set of information to the gateway. The gateway retrieves the requested information from CR and/or OCSP server, and responses the corresponding end-entity, as it is depicted in Fig. 5 and 6.

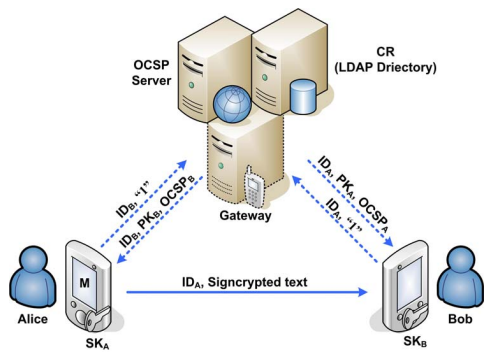

Figure 5. A typical message flow in the LPKI (mode "1": End-entities are capable of performing the required validations)

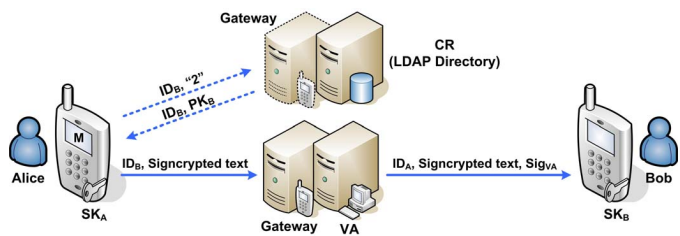

Figure 6. A typical message flow for the delegated configuration of LPKI (mode "2": All validations are delegated to the VA)

## V. CONCLUSIONS

In this paper, a *Lightweight Public Key Infrastructure* (LPKI) is introduced that is so suitable for the resource-constrained platforms, and for applications such as mobile commerce. It is based on the elliptic curve cryptography, deploys signcryption, assigns only one pair of private-public keys to each subscriber, delegates all the validations to a TTP called *Validation Authority* (VA). It has a compact compatibility with the well-tried PKIX infrastructure since its certificates have the same format of the popular X.509v3 certificates. However, this does not cause any problem since all the validations are delegated to the VA.